# Low-Resolution Imaging FMCW Lidar

Iskander Gazizov, Sergei Zenevich, and Alexander Rodin

*Abstract*—We demonstrate the imaging capability of a frequency modulated continuous wave lidar based on a fiber bundle. The lidar constructs velocity and range images for hard targets at a rate of 60 Hz. The sensing range is up to 30 m with 20 mW of output power. The instrument employs custom electronics with seven parallel heterodyne receivers. An example of image recovery is presented on 6-pixel "pictures" of a spinning disk and a drone hovering in the air. In experiments, we also tested the laser tuning linearity correction with a phase-locked loop. We see the practicality of such a low-resolution system as a boost in scanning rate of conventional lidars or for direct target imaging with a further upgrade of pixel count.

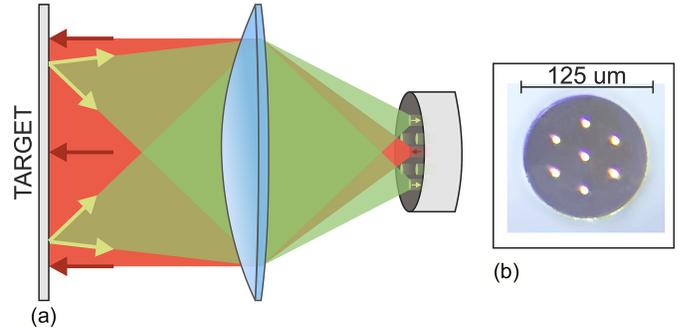

Fig. 1. (a) Lidar imaging configuration with fiber bundle and a lens. Emitted radiation from the central core (red) illuminates the target. Reflected radiation from different parts of a target (green) is focused on fiber bundle to be received by lateral cores. (b) Top view of 1 x 7 fiber bundle under a microscope.

## I. Introduction

TODAY, the demand for 3D mapping is reaching the degree when lidars are embedded in phones, autonomous vehicles, and robot vacuums [1]. The capability of measuring distances is often made possible with the time-of-flight commercial off-the-shelf (COTS) sensors through non-coherent detection [2]. Velocity measurements based on incoherent detection are widely applied for atmospheric wind measurements via Rayleigh/Mie/Raman and sodium lidars [3]. By contrast, coherent technology of frequency modulated continuous wave (FMCW) lidar is gaining attention owing to the benefits of simultaneous range and velocity measurements with high precision and low output optical power [4] and a downside of lower range, compared to other systems.

Apart from common trends, the application range of lidar technologies spans further, e.g., greenhouse gas remote sensing. Widely employed techniques include differential absorption lidars (DIAL) [5] and integrated path differential absorption lidars (IDPA) mounted on either ground-based [6] or airborne [7] platforms. Standard lidar-based laser infrared spectroscopy methods are widely applied onboard unmanned aerial vehicles (UAV) for a sounding range less than 100 m [8].

Historically, the first lidar applications were in meteorology [9] and military [10] more than half a century ago, shortly after the invention of the laser. Still, one could find a wide field for improvement in the niche of FMCW lidars: from advances in precision [11] and sensitivity [12] to the structures of phased arrays [13]–[15] and systems on chip [4], [16]. Furthermore, a common feature of most instrumental papers is diminishing laser chirp nonlinearity for boosting the sensitivity, which is done via preprocessing [17], [18] or in post [15], [19], [20]. There are various methods for chirp linearization during the measurement, including laser chirp predistortion [21] and real-time correction via phase-locked loops (PLL) [22].

One of the practical applications for lidars is 3D mapping, which could be implemented either by point-by-point scanning or with a matrix design for direct imaging. There is ongoing research around imaging lidars based on photon counting [23] and ghost imaging [24]. With the advances in chip manufacturing, it is now possible to produce direct time-of-flight (DTOF) sensors with a sub-megapixel resolution, orders of magnitude larger than FMCW imaging lidars [25].

Nevertheless, DTOF sensors lack velocity information, narrowing down their application area. Whereas imaging technology based on FMCW lidar provides information on both velocity and range, it is difficult to reproduce it even to a dozen channels with the added complexity of coherent detection [16]. One of the representative examples of an imaging FMCW lidar is an 8-channel device based on a photonic integrated circuit (PIC) with a maximum range of 60 m with only 5 mW of output power [4].

This work has been submitted to the IEEE for possible publication. Copyright may be transferred without notice, after which this version may no longer be accessible. Manuscript received May 13, 2022; revised XXX XX, 2022; accepted XXX XX, 2022. Date of publication XXX XX, 2022; date of current version XXX XX, 2022. This work was funded by the Russian Foundation for Fundamental Investigations under Grant 19-29-06104. (*Corresponding author: Alexander Rodin.*)

The authors are with the Moscow Institute of Physics and Technology (National Research University), Dolgoprudny, 141701, Russia, and with the Space Research Institute of the Russian Academy of Sciences, 117997 Moscow, Russia (e-mail: gazizov.ish@phystech.edu; zenevich.sg@phystech.edu; alexander.rodin@phystech.edu).

When constructing a receiver matrix for imaging FMCW lidar, each pixel is considered an independent coherent lidar accompanied by an optical scheme and broadband receiver. Thus, a lens array pointed at different parts of field-of-view is necessary for image construction, which in turn requires painstaking alignments. We propose a novel solution for imaging based on a *single* lens and a fiber bundle, a structure of multiple single-mode cores within a 125 μm diameter cladding in an FC/APC ferrule [26], [27]. The main advantage of the fiber bundle is the field of view segmentation, as different parts of the target image in the focal plane are received by different fiber cores (Fig. 1).

With the imaging capability, lidars could detect objects unreachable by radars [28]. For instance, the growing prevalence of small-scale drones in the air has been causing a variety of hazardous situations, including airport airspace violations [29]. This ubiquity of drones has recently aroused regulatory interest from government services. As a result, most countries have established rules for airspace access for UAVs, but compliance with these rules is limited by technical capabilities [30]. One of the tracking systems in development is the SafeShore complex, which applies lidars and conventional cameras for UAV detection [31]. Aircraft detection is also possible with the help of indirect signatures, including wake vortices [32] and sound correlation [33]. Thereby, the demand for drone detection in the civil field is growing along with the market expansion, and lidar could be the technology suited for future demands.

With the experience in lidar-based remote sensing instruments and multichannel spectrometers outlined in our previous projects [34]–[36], we propose a 6-pixel FMCW lidar as a proof of concept for low-resolution simultaneous velocity/range imaging. We further demonstrate the imaging capability with a fiber bundle on the examples of hard targets.

## II. METHODOLOGY

The idea behind the FMCW method lies in laser frequency modulation and onward processing of the frequency shift between emitted and reflected light waves. At every moment, there is a superposition of frequency shifts caused by the nonzero distance to the object and the Doppler effect in the case of object movement [20]. Since the frequency shift between emitted and reflected light is orders of magnitude lower than the laser's carrier frequency, it is preferable to operate with beat signals through heterodyne detection.

While modulating laser frequency with triangular chirps, beat signal frequencies $f_{up}$, $f_{down}$ are measured on rising and falling edges of the chirp. The detection method could be described by the equation (1), where $\nu_{Doppler}$ is frequency shift caused by the Doppler effect, $\alpha$ is laser frequency tuning speed in MHz per microsecond, and $\tau$ is the time delay caused by the distance to an object.

$$f_{beat} = |\nu_{Doppler} \pm \alpha\tau| \quad (1)$$

In the case of phase delay prevalence over Doppler shift, both the distance and speed of an object could be determined from (1) as:

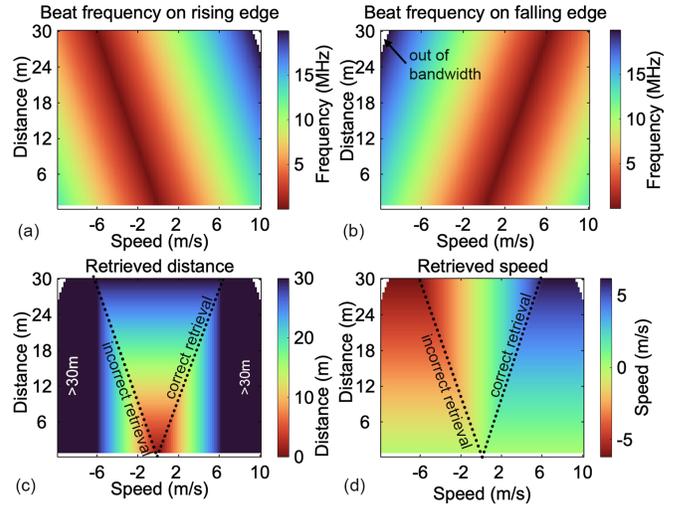

Fig. 2. Simulation of the beat frequencies received by lidar on falling (a) and rising (b) edges of laser chirp for various velocities and ranges. Further retrieval of distance (c) and speed (d) from the simulated set of beat frequencies demonstrates the problem of false retrieval.

$$d = \frac{c}{4\alpha}\left|f_{up} + f_{down}\right| \quad (2)$$

$$v = \frac{\lambda}{4} sgn\left(f_{up} - f_{down}\right)\left|f_{up} - f_{down}\right|. \quad (3)$$

However, the same equations output incorrect values when Doppler shift predominates over phase delay, as (3) would tend to zero in case of large speeds. This phenomenon is caused by the lack of knowledge of the signs inside of modules when transitioning from (1) to (2, 3). This problem could be readily resolved by swapping the signs inside the modules of (2, 3) or by introducing a complex frequency chirp, as described in [37].

To visualize the FMCW method, we present the distribution of beat frequencies depending on various target speeds and distances in Fig. 2(a, b), according to (1). In the simulation, the laser chirp range is 10 GHz with a 2 kHz repetition rate.

An inverse problem could also be visualized with the computed distribution of beat frequencies from Fig. 2(a, b). We retrieve both the distance and speed from (2, 3) as seen in Fig. 2(c, d). However, the problem of incorrect retrieval introduced earlier is noticeable, as only regions inside dotted lines are retrieved correctly. The other pair of equations should be applied for the regions with false speed/distance retrievals, which leads to uncertainty in the experiments with moving targets. This problem was resolved in other studies with a combination of frequency slopes [37].

## III. INSTRUMENT DESIGN

### A. Imaging System

The 1 x 7 fiber bundle by Chiral Photonics Inc. is responsible for the imaging capability of FMCW lidar. Fig. 1 demonstrates how the central fiber core illuminates the object, whereas the lateral cores receive reflected light. A single lens collimates the output beam and, at the same time, forms the image at the surface of the fiber bundle.

We simulated the imaging capability of the proposed scheme in Zemax OpticStudio to select the lens with the greatest differentiation between the individual channels. For best pixel separation, we focus the optical system on the target. The lens is chosen so that the minimal spot size on the surface of the fiber bundle from the point source is smaller than the pitch of the fiber bundle. Otherwise, all the pixels would represent the same values, and no image could be formed. The best optics in our stock appeared to be Thorlabs AL2520M aspheric lens with a 1-inch diameter and 20 mm focal length. This lens provides a 2 cm spatial resolution at a 20 m distance for a 35 μm pitch between fiber bundle cores. If we assume the sensitive area of the fiber bundle to be 80 μm in diameter, the total field of view of the lidar is 0.23 degrees.

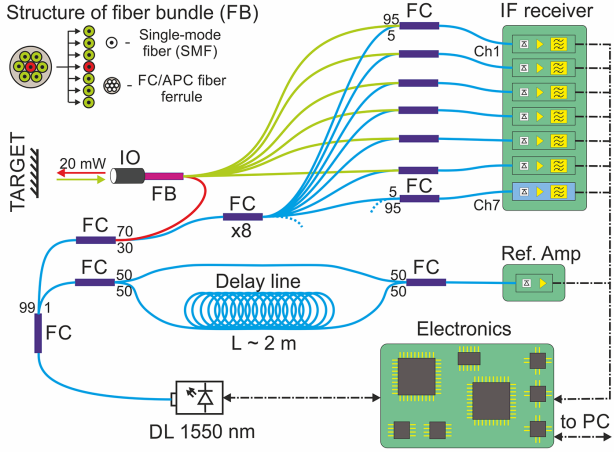

Fig. 3. Schematic of multichannel imaging FMCW lidar. DL is a diode laser, FC is a fiber coupler, IO is an imaging optics, FB is a fiber bundle, Ref. Amp is a reference amplifier, and IF is an intermediate frequency.

*B. Multichannel Design*

Several peculiarities should be considered when designing a lidar with multiple receiving channels. From the diagram in Fig. 3, radiation flux of 35 mW is split thrice: ~1% for frequency linearization, ~70% for target illumination, and ~30% as a local oscillator for heterodyne detection. The optical power of a local oscillator for each channel should be selected so that the overall signal exceeds the receiver noise to reach higher sensitivity. Therefore, an increasing number of channels requires an increase in laser output power. In our case, the laser emits 20 mW to free space. The fiber couplers (FC) ratios were selected experimentally to achieve the best sensitivity. We chose the Thorlabs SFL1550S single-frequency laser (SFL) for its high output power and narrow linewidth of 50 kHz.

Optical interference in the sophisticated fiber system is an undesirable consequence of the multichannel design, and it may eventually lead to false target detection. We introduced the seventh receiver channel disconnected from the imaging optics to eliminate interference in postprocessing (colored blue in Fig. 3). The observed interference pattern from this channel is subtracted from other channels in software to boost detection sensitivity.

Another complication of the multichannel design is parallel signal processing. At first, we designed a 7-channel amplifier board for signal amplification and filtering. Next, we developed a custom control board for parallel signal processing based on FPGA with multiple ADCs. The spectra are calculated on a PC with the LabVIEW software, as we ran out of FPGA memory when implementing seven FFT cores from Matlab HDL Coder inside an FPGA. Nonetheless, FFT in LabVIEW still provides a 60 Hz refresh rate for lidar image construction.

A custom laser driver board was developed to provide precise laser control. For laser linearity correction, we implemented a reference channel represented by an unbalanced Mach–Zehnder interferometer with a delay line of 2 m. The frequency of the beat note signal from the reference channel is proportional to the laser tuning linearity. A typical algorithm for laser linearity control is the phase-locked loop, described in detail in [17]. In a nutshell, the reference clock is generated inside an FPGA for XNOR operation with the digitized beat signal from the reference channel amplifier. The result of XNOR is accumulated throughout the rising and falling edges of the laser chirp and fed to laser current DAC in parallel with the main laser chirp. Notably, the laser's DAC sampling rate is several times higher than the main laser chirp sampling rate, allowing the PLL reference clock period to be selected independently from the sampling rate of the main laser chirp. The delay line length is chosen so that the beat frequency is in the kHz range to resolve the beat note signal.

*Lidar Frequency Tuning*

The main idea of the FMCW method is the modulation of laser frequency and further analysis of the frequency-shifted reflected signal. In our case, a basic triangular laser chirp is employed. The tuning range of the SFL laser is technically limited by 5 GHz, but it was reduced to 3 GHz because of excessive interference observed in the signal. We noticed that multiple fiber connectors induce the interference pattern, and crosstalk in fiber bundle could also supplement to the cause.

With a modulation period of 1 ms, the frequency sweeping rate $\alpha$ is 3 MHz/μs. The IF receiver bandwidth is 20 MHz, which results in a frequency resolution of 19 kHz, considering the 2048-point FFT. By differentiating (1), one may find that the corresponding spatial resolution for range-finding is 0.5 m, and 0.015 m/s for velocity retrieval. Quite a rough distance accuracy is a consequence of the narrow sweeping range of the SFL laser. Nevertheless, the accuracy could be improved by exploiting a laser with narrower linewidth and a frequency sweep in the THz range [38].

We explored a phase-locked loop (PLL) algorithm for laser sweeping linearity stabilization by measuring the distance to a wall. As from the works [17], [18], [21], it is common to implement initial passive predistortion algorithms for linearization with the active PLL over it. In the scope of this work, only PLL algorithm is applied for simplicity, as the frequency sweep range of an SFL laser is narrow enough. Though, the lack of predistortion could have affected the final results.

We demonstrate the algorithm output in Fig. 4(a, b) with a 50 ms sweeping period for easier visualization. The field testing of the algorithm was performed with a 1 ms sweeping period in Fig. 4(c, d). The laser current shape distortion in Fig. 4(a) is noticeable in the algorithm's attempt at linear frequency tuning.

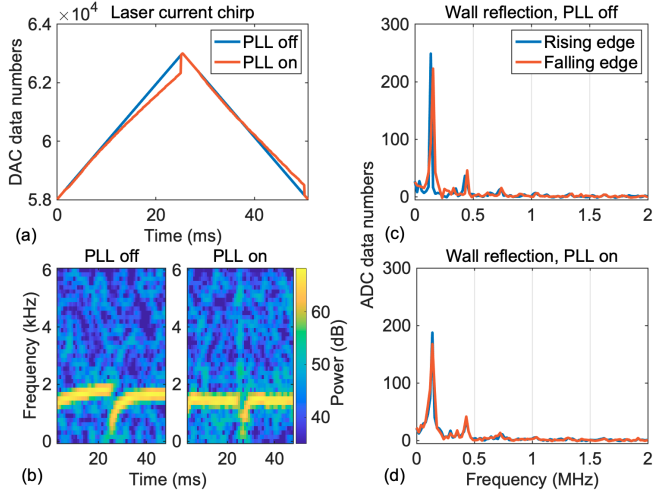

Fig. 4. (a) Laser current chirp shape when the phase-locked loop (PLL) algorithm for frequency sweep linearization is turned on (red) or off (blue). (b) Measured spectrogram of the beat signal in the delay line interferometer when PLL is turned off or on. Spectra of the signal reflected from the wall at ~7 m with PLL turned off (c) or on (d).

The spectrogram of the beat signal from the delay line interferometer visualizes the difference between linear and nonlinear laser frequency tuning in Fig. 4(b). The discreteness of spectrograms results from data transfer optimization from lidar to PC.

Laser frequency sweep linearization aims to boost lidar measurements' sensitivity and accuracy, as the nonlinear tuning results in broadened received beat note [39]. We observed the signal reflected from the wall at ~7 m with a 1 ms sweeping period to test the algorithm. As seen in Fig. 4(c, d), the signal amplitude decreased with PLL enabled. The amplitude drop could be explained by the decrease in frequency sweep range noticeable in Fig. 4(a), which led to spatial resolution reduction. We reproduced the same effect of amplitude drop by intentionally reducing the frequency tuning range.

Nevertheless, the frequency shift between the spectra in Fig. 4(d) related to rising and falling edges was eliminated with enabled PLL, which improved the accuracy of speed and distance retrieval.

## IV. RESULTS

### A. Observation of a Disk

We carried out initial attempts at speed/distance image acquisition with a white painted compact disk mounted on a brushless motor in front of the wall. This setup provided both moving and stationary targets, as noted on spectra in Fig. 5. After spectra acquisition, the software subtracted the interference channel from the imaging channels to increase signal detectability. Next, the software retrieved the positions of the highest peaks for every pixel (imaging channel) for range and velocity calculation according to (2). From calculated speed/distance values, software constructed lidar images at a rate of 60 Hz.

Finally, we combined lidar images into a multipixel mosaic by manually sliding the laser spot across the disk, as shown in Fig. 6. Sharp edges of the disk resulted in a noticeable contrast of the "images." The Xenix XEVA-3397 InGaAs camera took photos of the laser spot on the disk. The distance to the disk was 3.5 m and 7 m to the wall. With this experiment, the distance accuracy was ±0.4 m, and the axial speed accuracy was ±0.02 m/s.

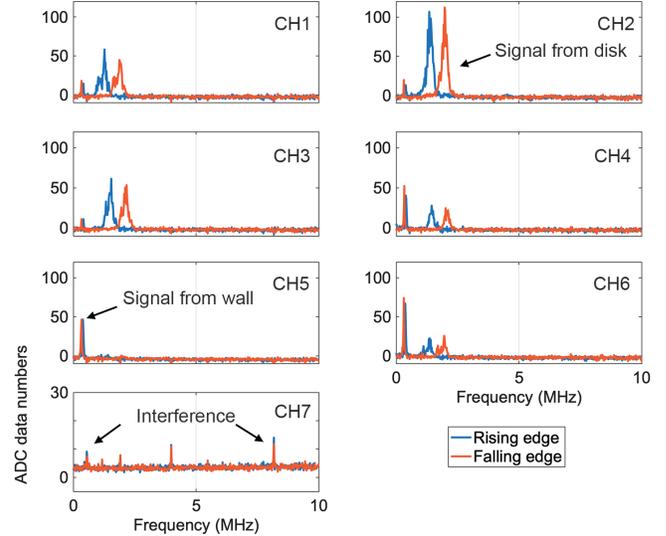

Fig. 5. Spectra from all seven lidar channels in the experiment where lidar illuminates a rotating disk with a wall as a background. The first six channels (pixels) represent different fiber bundle cores. There is a noticeable differentiation between signals from channels, as every channel receives the reflected light from various regions of the lens field-of-view. The seventh channel is for interference pattern observation only. The first six spectra correspond to lidar image #1 in Fig. 6.

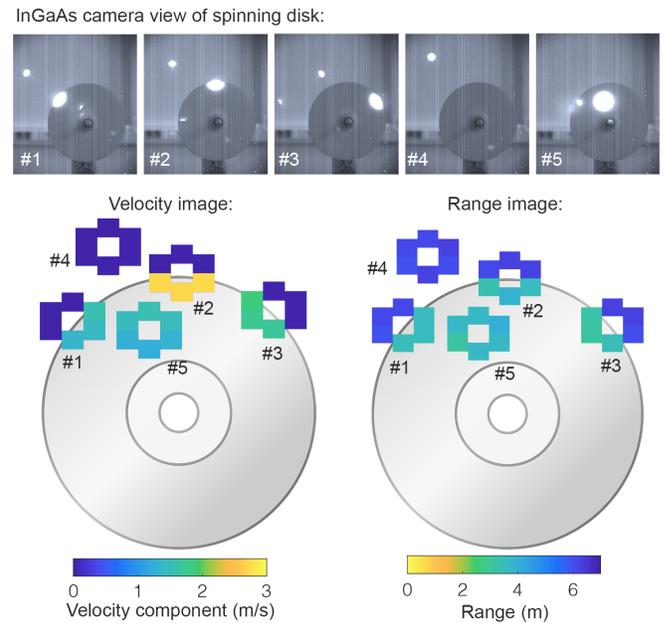

Fig.6. Top: images of the laser spot on different parts of rotating disk taken with InGaAs camera. Bottom: resulting mosaic of velocity and range images from the lidar. The indexes on InGaAs images correspond to pixel indexes on lidar images. The scale of lidar images is selected for visualization purposes and does not represent the actual field of view.

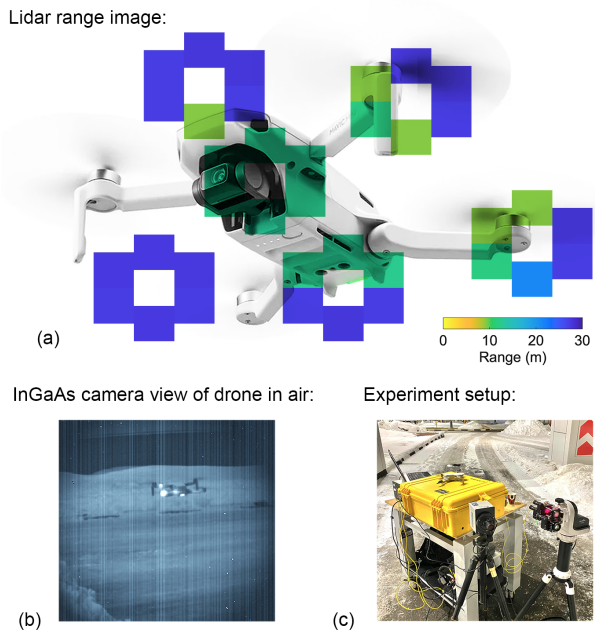

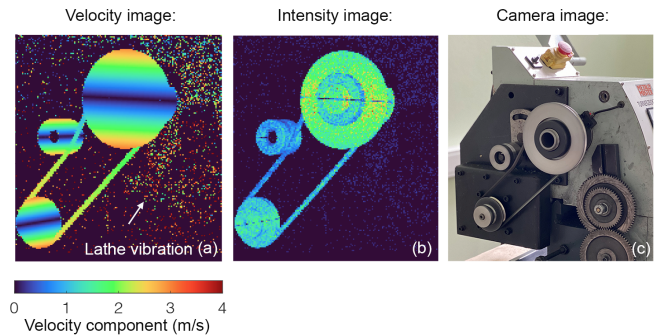

Fig. 7. (a) Resulting mosaic of distance images in the experiment with a drone hovering 15 m from lidar. The scale of lidar images is selected for visualization purposes and does not represent the actual field of view. (b) Image of the laser spot on the drone from InGaAs camera. (c) Experiment setup with the lidar packed into the yellow case.

Fig. 8. (a) Image of lathe velocity projection to the line of sight which was scanned with single channel of our lidar. We employed the laser Doppler velocimetry technique to construct this image. (b) Complementary image of beat note amplitudes. (c) The running lathe which was scanned point-by-point with lidar on telescope mount.

### B. Observation of a Drone

As proof of concept for drone detection, we imaged the DJI Mavic Mini while in the air from 15 meters. Fig. 7(a) represents the resulting mosaic of distance images overlaid on a drone for illustration purposes. The bright lidar laser beam on a drone in Fig. 7(c) was observed by an InGaAs camera for beam guidance when constructing a mosaic. During these outdoor experiments, necessary safety precautions were taken.

Obtained mosaic proved the ability of the proposed lidar with fiber bundle to construct images of real-world targets outside of the laboratory environment. All conducted experiments demonstrated correct results, as the distance results were validated via laser rangefinder and velocimetry data with a tachometer.

## V. Discussion

We presented the imaging lidar capable of producing 6-pixel pictures of range and velocity in real-time. Parallel heterodyne receivers acquire the spectra for every pixel from the fiber bundle to construct an image. Signal preprocessing and experiment control are implemented in FPGA. In experiments, we successfully constructed distance and speed images of rotating disk and drone hovering in the air. The algorithm for laser frequency tuning linearization was also implemented and tested.

We designed this instrument solely as a proof of concept of an imaging system based on a fiber bundle. Yet, there are directions for further improvements, and we see the practicality in real-world applications. The first option is a 1 x 22 fiber bundle to produce higher resolution images. This option could dampen the speckle phase noise for a non-scanning sounding [40]. Particle velocimetry might also benefit from such an imaging system, either for trajectory reconstruction or sensitivity improvement, depending on imaging optics. The second option is with the scanning system applied to the current design, increasing the scanning rate six times as of traditional single-channel lidar. We tested scanning mode on a single lidar channel, as seen in Fig. 8. And we think that emerging developments in laser Doppler vibrometry might benefit from such an application of a fiber bundle [41].

### Acknowledgment

The authors would like to thank Oleg Benderov for advice during the experiments, as the core idea of applying fiber bundle belongs to him. We thank Shamil Gazizov for the help in the development of mechanical test benches. Finally, we thank Maxim Spiridonov, as only with his experience we were able to develop the electronics of the lidar.

Authors' biographies not available at the time of publication.